\documentclass[aps,pre,twocolumn]{revtex4}

\usepackage[utf8]{inputenc}
\usepackage[dvips]{graphicx}

\graphicspath{{./data/}}

\begin{document}


\title{Vertebrate pollinators: phase transition in a time-dependent generalized
traveling-salesperson problem}

\author{M. Jungsbluth}
\affiliation{The unbelievable Machine Company GmbH, Berlin, Germany}
\author{J. Thiele} 
\affiliation{University of T\"ubingen, Germany}
\author{Y. Winter}
\affiliation{Institute of Biology,  Humboldt University 
and Center of Excellence  NeuroCure, Berlin, Germany}
\author{H. Schawe}
\affiliation{ Institute of Physics, 
   University of Oldenburg, Germany}
\author{A. K. Hartmann}
\email{a.hartmann@uni-oldenburg.de}
\affiliation{ Institute of Physics, 
   University of Oldenburg, Germany}

\date{\today}

\begin{abstract}

We introduce a model for the
global optimization problem of  nectar harvesting by flower visitors,  
e.g., nectar-feeding bats, as a
generalization of the (multiple) traveling-salesperson problem (TSP). 
The model includes multiple independent
 animals and many flowers with time-dependent content.
 This provides an ensemble of realistic combinatorial optimization
problems, in contrast to previously studied models like 
random Satisfiability or standard TSP.

We numerically studied the optimum harvesting of these foragers,
with parameters obtained from experiments, by  using 
genetic algorithms. For the distribution of travel distances, 
we find a power-law (or L\'evy) distribution, 
as often found for natural foragers.
Note, in contrast to many models, we make no assumption about the
nature of the flight-distance distribution, the power law just emerges.
This is in contrast to the TSP, where we find in the present study 
an exponential tail.

Furthermore, the optimization problem exhibits a {phase transition}, 
similar to the TSP,
at a critical value for the amount of nectar which can be harvested. 
 This phase transition coincides with a dramatic increase
in the typical running time of the optimization algorithm. For the 
value of the critical exponent $\nu$, describing the divergence
of the correlation length, we find $\nu=1.7(4)$, which is on the other
hand compatible with the value found for the TSP.

Finally, we also present data from field experiments in Costa Rica
for the resource use for freely visiting flower bats. We found that
 the temporal patterns in experiments and model agree  remarkably,
confirming our model. Also the data show that the bats are able to
memorize the positions of food sources and optimize, at least partially,
 their routes.
\end{abstract}


\maketitle



\section{Introduction}

Evolution is a process of the constant optimizing of a species
regarding its survival and reproduction capabilities. One particular
activity each animal has to perform is harvesting 
\cite{stephens2007,prins2008}. 
 Hence nature  poses another optimization problem, since
every single animal has to optimize its own effort in  obtaining as much
food as needed with as little effort as possible in a potentially dangerous 
world.

However, each animal competes with other animals of the
same and other species. Thus the food resources available within a
given area are harvested simultaneously by, in our case, multiple bats. This
poses the question of whether the single-animal optimization can
even take into account the activity of other bats, such that
overall optimal or near-optimal behavior is reached by 
the total group of animals. Being able to calculate such overall
 optima would allow the determination of how the efficiency of  a
bat's behavior approaches the 
theoretical optimum solution. 
To date, a theoretical approach for
this everyday animal foraging problem of determining minimum-effort
harvesting optima in multiple bat and time-dynamic resource
landscapes has, to our knowledge, not been published. 
With this study we present a
corresponding model and an approach to determining such optima.
Note that for static non-time dependent situations, simulations to
detect optimal harvesting exist, e.g., for the social foraging
behavior of bacteria like E.coli \cite{liu2002}. On the other hand,
also many models for search in random environments, but without
spatio-temporal memory exist \cite{viswanathan1996,van_dartel2009,wosniack2017}.

From the theoretical point of view,
optimization problems have been moved into the focus
of statistical physics, where the collective behavior of many-particle
systems is investigated. Of particular interest are phase transitions,
which are abrupt changes of  the properties of a system, e.g., for the
solid-liquid transition of water when increasing the temperature above
the melting point.
Few years ago,
phase transitions have been observed  
\cite{phase-transitions2005,mitchell1992,gent1996,moore2011} 
also for simple random ensembles of optimization problems,
like the  (TSP), K-Satisfiability (K-SAT), 
or the vertex-cover problem. 
These ensembles have the advantage of 
being easy to describe but among practitioners they are often 
criticized as being very artificial and far from real-world problems.

Therefore, we study here an optimization problem directly occurring in nature,
the problem \cite{papadimitriou2000,phase-transitions2005} 
of nectar harvesting by flower visitors. Nectar harvesting behavior 
leads to a single movement path during daily activity, with intermittent 
stops at flowers to collect nectar. The length of this path can 
serve as a measure of the cost of harvesting. 
This means we want to find spatio-temporal sets of paths, 
one path for each animal, such that the overall net gain of energy is 
maximized. The global optimization of the ``nectar-harvesting problem'' (NHP) 
is  a generalization of the (multiple) traveling-salesperson problem (TSP) 
\cite{gutin2002}, 
which is known to be \emph{NP-hard} \cite{garey1979}. This means,
that the NHP is computationally hard to solve and the only algorithms
available are those, where the worst-case running time grows
exponentially with the size of the problem
(i.e.\ the number of flowers available in  the environment).  
In comparison with the standard TSP there are further special 
characteristics of foraging at replenishing resource locations. First, a 
flower's value is not only 0 or 1 but can take intermediate values 
depending on its current nectar content. 
Furthermore, flowers are revisited within the daily activity period as 
nectar is replenished through ongoing secretion. Thus we can
expect that, from the computational point of view,
 the nectar harvesting problem NHP is even harder to solve than the TSP.

Note that the NHP does not take into account if a neural system could 
actually implement the optimization process used, instead
it should serve as a null-model for the best optimizing
bats possible, to which real bats and models based on learning theory
can and could be compared with. Nevertheless, it is well known
that swarm intelligence may
lead to finding global optima. This is used in nature-inspired
approaches like \emph{ant-colony optimization} \cite{dorigo2004,pintea2014}. 
Therefore, investigating
(near) global optima of the NHP might tell actually a lot about the global
result of the individual 
behavior of real bats, even tough in the moment the capabilities
of the individuals with respect to optimization are yet not extensively known.

The notion that bats optimize their behavior was motivated
by data from field experiments in the Costa Rican rain forest, where 
we investigated the spatio-temporal behavior of 
nectar-feeding bats \cite{winter2001,helversen2003}, 
which exploit resource landscapes with 
flowers at fixed locations offering nectar to their visitors in 
variable amounts depending on secretion rate and past exploitation 
events. Flower fields thus constitute potentially predictable resource 
distributions with deterministic spatio-temporal dynamics. Vertebrate 
flower visitors with cognitive abilities may have evolved mechanisms to 
adapt their resource-exploitation behavior to such dynamics.
Hence each animal may be able to optimize its own
behavior, justifying the comparison with the result of an optimization
approach.

We analyzed the step-size distribution of the optimum paths.
We observed a power-law, i.e., L\'evy distribution of the flight distances
as an emergent property of our model, without assuming a power-law
distribution anywhere in the model. Note that for the related but much
more abstract and non-time-dependent 
traveling-salesperson problem (TSP), our results presented 
below do not show power-law
behavior but a faster exponential decrease.
For the NHP, the power-law nature of the distribution
is a true emergent property of the system and its optimization.
 Such distributions have been observed often in nature 
\cite{cole1995,viswanathan1996,humphries2010,raichlen2014}
and used, i.e., assumed in modelling approaches 
\cite{viswanathan1999,reynolds2009,gautestad2013,zhao2015,kusmierz2017}.
It is an open question \cite{bartumeus2007} 
whether the L\'evy-type distributions seen in natural systems arise
 from an evolutionary process,
as seen in some models \cite{van_dartel2009,wosniack2017}, 
or whether it is an emergent property of the
system, as for the present case.

As we will furthermore show below, the NHP also exhibits a phase transition
in terms of solvability, and computational complexity.
This is similar to, e.g., K-SAT or TSP 
 but in contrast
to these previously studied model the NHP describes a quite realistic
ensemble of real-world optimization problems.
With our experiments we tested the validity of the three central 
assumptions on which our theoretical analyses are based and gathered 
evidence that real bats indeed optimize their foraging behavior. We 
assume that a forager discriminates within patches between resource 
locations differing in profitability. This requires spatial reference 
memory. Furthermore, a forager must remember its own foraging actions 
in order to avoid otherwise profitable locations after depletion and 
before subsequent replenishment. This requires spatial working memory. 
Thirdly, a forager able to visit resource locations along shortest 
travel paths needs besides a cognitive spatial representation the 
ability for odometry to discriminate path lengths. Our experimental 
data support these assumptions. Furthermore, our method of examining 
decision making and foraging behavior with free-ranging animals 
provides a novel experimental approach of how to actually test 
predictions of theoretical models under natural conditions. 

\section{Material and methods}

\subsection{Model}

An animal's daily foraging tour consisted of a series of visits to 
flowers. Food intake during the visit to a flower depended 
on the current nectar content of the 
flower at the time of a visit, and was limited by 
the bat's maximal intake rate  
and by its stomach capacity. Each tour was characterized by a 
total amount of nectar intake and a total amount of net energy gain,
determined after the energy cost of foraging activities was 
subtracted.

A habitat was modeled as $M$ bats living in a square area of size one hectare
($10^4$m$^2$),
where randomly placed flowers $k=1,\ldots, N$ are available for feeding. 
Our model and the parameters we used are based on experimental data
for bats  obtained in earlier studies 
\cite{winter2001,helversen2003,winter1999}.
We studied habitats with up to 24 bats having 10 to 120 flowers available.
 The positions
of the flowers determine the matrix
of distances $d(k,k')$ between flowers $k,k'$.
Each flower $k$ 
produces nectar with rate $p_k(t)$, $t$ is the time during one night,
in which at $t=0$ h the bats start harvesting.
 Here we considered
usually constant production rates $p_k(t)=$ const $=50.4\,\mu$l/h for 
$t\in[0,12]$ h. 
For some simulations we also used triangle-like rates, which have 
$p_k(-1.5h)=p_k(8.5h)=0$ and rise from $t=-1.5$ h linearly with time until
$p_k(0.5h)=p_{\max}=57.6\,\mu$l/h, and decrease linearly with time afterward
until $t=8.5$ h.
Each flower can hold a maximum amount $C_k$ of nectar, here $C_k =200\mu$l

The behavior of each bat $j\in \{1,\ldots,M\}$ 
during one night is described by a series
of $N^{(j)}$ visited 
flowers $s_i^{(j)}\in \{1,\ldots,N\}$ ($i=1,\ldots, N^{(j)}$)
 and time moments $t_i^{(j)}$ when the flowers are reached. We assumed that all
bats fly with constant speed $v=3$m/s. Hence, the time to travel between
the $i$'th and $(i+1)$'th visit is 
$\Delta t_{i+1}^{(j)}=vd(s_i^{(j)},s_{i+1}^{(j)})$. The remaining time 
$t_{i+1}^{(j)}-t_i^{(j)}-\Delta t_{i+1}^{(j)}$ a bat spends  
at the $i$'th visit, most of the time at rest.
 An ``instance'' is  a completed simulation based on a 
set of (randomly chosen) 
\emph{positions} of the
flowers together with resulting  travel \emph{schedules} 
$x=[(s_1^{(1)},t_1^{(1)}),\ldots,(s_{N^{(1)}}^{(1)},t_{N^{(1)}}^{(1)}),
(s_1^{(2)},t_1^{(2)}),\ldots]$
of all $M$ bats during one night.

We are interested in the net energy balance of an instance $x$. 
 The net energy balance $f(x)$
is the harvested nectar minus the nectar expended by the bats for
the energetic cost of their tours (all energies
are measured in terms of amount $\mu$l of nectar). 
To be more specific, the
 net energy gain per bat  is given by
\begin{eqnarray}
  f (x) & =  & \frac{1}{M} \sum_{j = 1}^M \Big(
\underbrace{E^{(j)}_{\rm collect}(x)}_{\textnormal{time-dependent}} - 
\underbrace{E^{(j)}_{\rm travelcost}(x)}_{\textnormal{position-dependent}} 
\nonumber\\
& & 
- \underbrace{E^{(j)}_{\rm stopcost}(x)}_{\textnormal{stop-dependent}}\Big).
\label{eq:score_nodee}
\end{eqnarray}
\begin{itemize}
\item
$ E^{(j)}_{\rm collect}(x)$ is the amount of nectar bat $j$ collects. If a
bat  arrives at a flower, it takes as much nectar as possible, i.e.,
limited by the amount of available nectar at the given time (which is 
determined by the nectar produced so far and  the nectar taken during
previous stops by this and other bats) and by the capacity $C^{(j)}$ 
of the stomach, here  $C^{(j)}=500\,\mu$l for all bats.
Nectar is digested with rate $r_{\rm digest}$, hence
after some time a full stomach can take additional nectar again.

\item
$E^{(j)}_{\rm travelcost}(x)$ determines the energy consumed by flying between
the different visits and is determined by the total flight distance
and by the energy consumption $c_{\rm length}=0.4\,\mu$l/s  
(with velocity $v=3\,$m/s).

\item 
$E^{(j)}_{\rm stopcost}(x)$ accounts for a higher energy need when 
feeding at a flower
(hovering flight) and is, for simplicity, for each bat $j$ 
just a constant energy $c_{\rm stop}=1.2\mu\,$l
times the number of stops $N^{(j)}$.
\end{itemize}

For our simulations, we are interested in maximizing the net energy gain
$f(x)$ for a given distribution of the flowers and number of bats.
 It is the net gain that an animal can invest directly or indirectly into reproduction.  
We used a genetic algorithm briefly described in the next subsection. This 
energy gain increases when the cost for collecting the nectar decreases.
Biologically speaking, this would mean that a bat 
minimizes its own locomotion effort and that it also minimizes 
the effects of exploitative competition between itself and the other bats.
In particular, we want to determine
whether under given ecological conditions 
it is possible in principle for each bat to obtain enough
nectar to meet at least  its basic requirements. Having the
optimum solution available, it is then possible to compare it with 
the performance of
 real bats. Agent-based simulations could also be used to test 
assumptions about information gathering, information processing, 
and decision making by individual or competing bats, extending
previous experimental laboratory studies on the spatial memory 
and spatial cognition of
bats \cite{thiele2005,winter2005}

\subsection{Simulation Algorithms}
\label{sec:methods}
We have performed computer simulations \cite{practical_guide2015}
using genetic algorithms \cite{goldberg1989} and using a greedy algorithm 
(see below) to find optimum or at least near-optimum solutions.  

The basic idea of an evolutionary (or genetic) algorithm is to
 mimic the evolution of a group of possible schedules
 (often called ``individuals'' in this context,
 not to be confused with the bats) to the same problem.  
Schedules that adapt
 better to the optimization task (i.e.\ have a higher ``fitness'')
 have a higher probability of survival. Thus they pass their genes in the simulation process more frequently to subsequent generations than others. This means that
 the average fitness of the population of schedules
  increases with time, hence one
 gradually approaches the optimum schedule.  To be specific, we used a
 ``population size'' (not to be confused with the number $M$ of bats)
 ranging from 100 schedules $x^{(z)}$ (smallest system size) to
 1000 realizations,  tournament selection
 \cite{goldberg1991,blickle1995},  and an adaptive mutation rate of
 $1/N^{(j)}$ for each bat. Mutation operators allowed the random
 changing of the
 time a bat pauses after a series of visits  
by a small amount,  as well as
 the  insertion/deletion/exchange of one or several flowers within and between
 different schedules of an instance.  As a crossover mechanism 
we took ``time-point
 crossover'', which means that a new schedule $x^{\rm new}$  can be
 built from taking the sequence of flowers visited before some
 absolute (randomly chosen) time $t_{\rm cross}$  from one schedule
 $x^{(z_1)}$ (for all bats),  and the sequence of flowers visited
 after  $t_{\rm cross}$ from a second schedule $x^{(z_2)}$.  We also
 applied local optimization steps, where changes are only accepted if they
 increase $f(x)$. Possible changes here are  replacing a visited
 flower by another flower, changing the order in which two flowers are
 visited, or  changing the time a bat pauses at a flower. 
 To find the global optimum
 schedules with high probability we applied up to $10^5$
 generations of the genetic algorithm.

We also computed exact solutions for single 
TSP problems \cite{bektas2003}. We used this partially 
to generate initial schedules 
for the genetic algorithms, but also for the purpose of comparison
of some quantities. Technically, we used the CONCORDE library.
\cite{applegate2003}.

Also a ``greedy'' algorithm was used to find a solution 
of the NHP. Solutions from this algorithm may not converge as closely to
 the optimal solution as with the genetic algorithm. However, we applied it as the greedy heuristic derived from reinforcement learning could actually 
be similar to a strategy used by real bats.
 The greedy algorithm did not use the mutation and recombination 
methods of the genetic algorithm but instead 
proceeded as follows:
\begin{itemize}
\item each bat approaches the flower with highest current energy gain  
(energy intake - energy costs for flight)
\item if the stomach of the bat is full, rest until fully digested
\item if highest energy gain is negative (cost of flying is higher  
than any nectar content) wait for 10 minutes and try again until a  
positive best energy gain is found.
\end{itemize}
As in the genetic algorithm, several problem instances were  
calculated with this strategy: typically an average over 800 runs was 
performed. Note that these systems are computationally easier, hence
we could study larger systems.

\subsection{Experiments}

\begin{figure}[!h]
        \centering
\includegraphics[width=0.95\columnwidth]{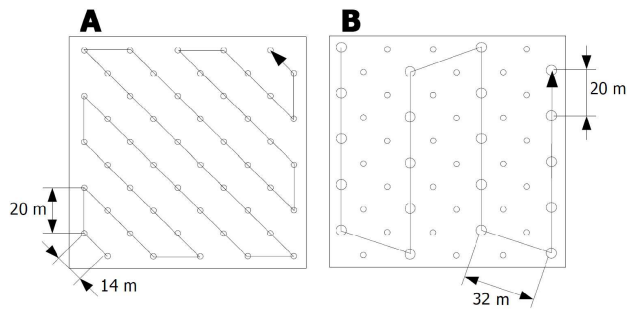}
\vspace*{3mm}
        \caption{Experimental 100x100 m flower field in the Costa Rican rain forest. Circles in A, B show positions of artificial flowers that were either quickly refilling QR (large) or slowly refilling SR (small). Minimum distance between neighbors was 14 m while minimum distance between QR type flowers was 20 m. Lines in A, B show hypothetical minimal travel paths for a forager that does not discriminate between flower qualities (A) and for a forager only visiting QR flowers (B). Note that flower qualities differed as shown in (B) throughout the experiment.
\label{fig:experiments}}
\end{figure}

Experiments were conducted to gather evidence that real bats indeed optimize flight behavior.
Data were obtained from eight individuals of free-flying 
Glossophaga commissarisi bats visiting a field of 50 artificial flowers \cite{thiele2006} in the rain forest  at La Selva Biological Station, Costa
 Rica (Fig.\ \ref{fig:experiments}).

Artificial flowers were computer
 controlled and produced nectar at a quality and rate found in natural
 flowers.
Ongoing secretion occurred into a virtual nectar account that determined nectar availability during an actual visit by a bat.
 Bats had been caught and
 equipped with passive transponder identification tags (RFID PIT tags) prior to the
 experiments. 
We present the data from the first three days after onset of the experiment. Nectar-feeding bat species other than Glossophaga commissarisi were uncommon on the experimental plot ($<5\%$ captures) and did not participate in the experiment. 

Each artificial flower was equipped with an electronic
 reader that identified a bat in real-time as it visited
 a flower. This allowed counting  on an individual basis
the visitation data of the
 free-flying bats.

 Also, the delivery of nectar was dependent on the individual identity of the visiting bat. This allowed us to operate the flower field in a ``non-competition'' mode where separate sets of virtual nectar accounts that were kept for each individual allowed depletion and replenishment of nectar to be made individually specific and independent from the potential competition by others. This was useful, since we wanted to investigate cognitive abilities of individual bats by exposing them to experimentally defined conditions of food availability that remained unaffected by other individuals. Our artificial flowers were of two types ``slowly refilling'' (SR, 30 mins to refill) and ``quickly refilling'' 
(QR, 15 mins to refill). Flowers were arranged such that the closest neighbors of QR flowers were all SR (diagonal direction, 14 m) while the next QR type was further away (up-down direction, 20 m). The rationale behind this experiment was that while a forager na\"ive about flower qualities should minimize path length by flying diagonally, a forager with knowledge of flower quality should fly in up-down direction predominantly visiting QR flowers. Thus we expected a forager that optimizes to have a preferred angle of traveling within the flower field and that this angle would change, assuming quality is discriminated, after the spatial positions of QR flowers were learned. 

\newpage 

\section{Results}

Below we state the results of our simulations to find optimized
joint exploitations of a habitat with randomly arranged flowers. 
The distribution of travel distances exhibits a power law as often found
for natural systems. Also,
we find a second order phase transition between a phase where enough resources
are available to a phase where this is not the case. We characterize the
phase transitions by means of finite-size scaling.

Also, we show results from the experiments in the rain forest, which,
considering the frequency of visits during the night, agree qualitatively
with the simulation results. Furthermore, the results indicate that the
bats use their spatio-temporal memory to optimize their foraging paths.

\subsection{Properties of optimized schedules}

\begin{figure}[!htb]
        \centering
\includegraphics[width=0.95\columnwidth]{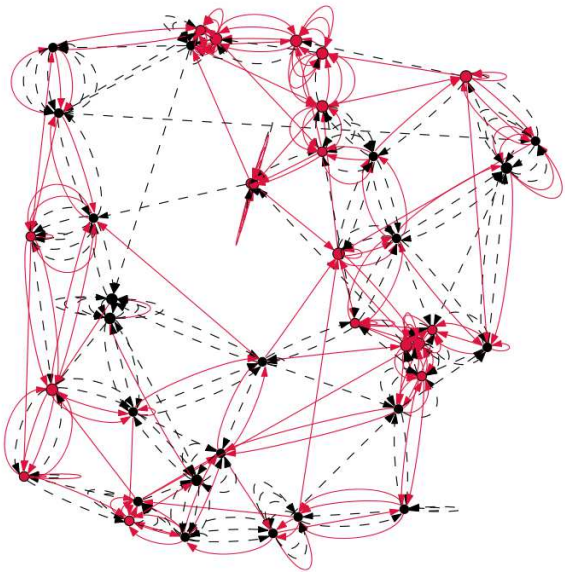}
        \caption{(color online) 
Example of an optimal solution obtained for two bats
and  40 flowers. The path of one bat is marked by full lines (red), the path
of the other bat by broken lines (black).}
        \label{fig:sample_tour}
\end{figure}

An example for an optimum solution of a one-night 
schedule for two bats having available
40 flowers is shown in Fig.~\ref{fig:sample_tour}. One can observe that
all flowers are visited several times, but not too often, since it is
beneficial after emptying a flower to wait before the flower
is revisited. Correspondingly, the bats spent most of the time
waiting at some flowers, which costs the least amount of energy.
Note that for this sample solution, both bats visit basically all flowers.

\begin{figure}[!t]
        \centering
\includegraphics[width=0.95\columnwidth]{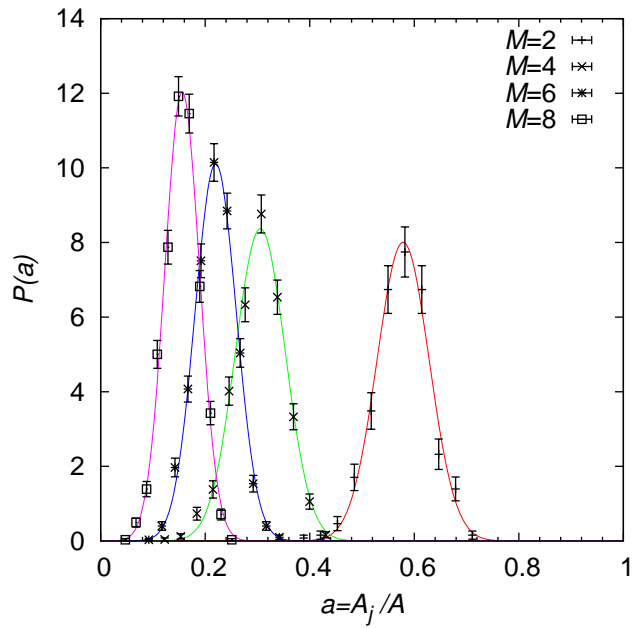}
        \caption{(color online) 
Distribution of area coverage for 80 flowers and 
different numbers $M$ of bats, averaged over 200 random instances.
The lines are guide to the eyes only, obtained by fitting to Gaussians. }
        \label{fig:area_coverage}
\end{figure}

Nevertheless, for larger number of bats and larger number of flowers,
it becomes beneficial that the bats partition the area among them, because
this reduces the overall flight distance slightly. To quantify
this, we have measured the ``area coverage'' which is defined as follows:
To each flower $s$ we assign a circle $C_{s}$, 
centered at the position of the
flower, which has a radius $r=\sqrt{\frac{A}{2N}}$ and therefore 
an area $\pi A/(2N)$.
 For each bat $j$ of a tour $x$,
we take the union of all circles for the flowers visited and
measure its covered area, i.e.,
\begin{equation}
A_j=\left|\bigcup_i C_{s_{i}^{(j)}} \right|
\end{equation}
Therefore, flowers which are close to each other will not contribute so much
to $A_j$, because the corresponding circles overlap extensively. 
In this way the quantity $A_j$ takes into account the geographical
distribution of the flowers, as compared to measuring just the fraction
of visited flowers.
This makes sense, because visiting close-by flowers will not cost so much 
energy for the bats.
The measurement is done each time by two-dimensional numerical integration.
In Fig.~\ref{fig:area_coverage}, the distribution of the 
relative area coverage $a=A_j/A$  (for all bats $j$), where
$A$ is the total area, is shown for different number of bats and
always 80 flowers. Note that the relative coverage of a bat visiting 80
randomly distributed flowers has a mean of approximately $a=0.76$. 
Therefore, already 2 bats do not visit during a night all
flowers, but almost all. Furthermore,
one can
observe that $a$ decreases on average with growing number of bats. This means,
with increased competition, it becomes, to achieve a global
cooperative optimum,  more beneficial to partition
the area among the bats. Therefore, only through the global
optimization constraint, an indirect repulsive interaction among the
bats is introduced.

\begin{figure}[!t]
        \centering
\includegraphics[width=0.95\columnwidth]{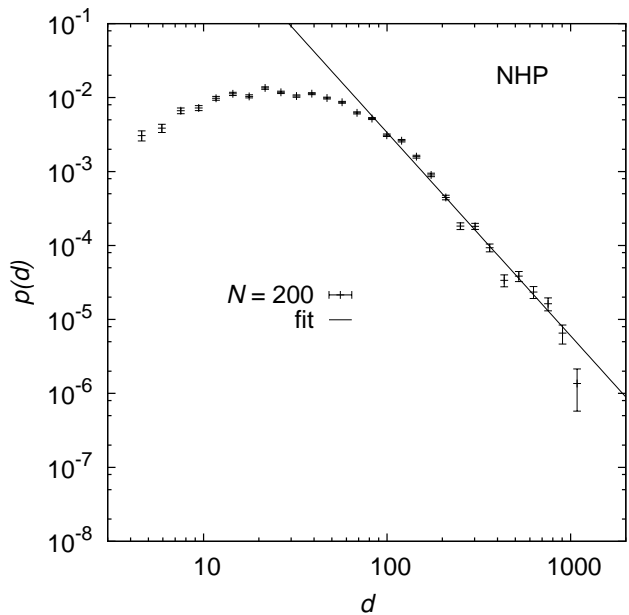}
        \caption{Distribution of travel distances $d$ between two stops
shown on log-log scale, here for $N=200$ flowers, two bats on 
an area of $A=10^6\,$m$^2$, averaged over
5000 instances.
The line indicates a fit of the tail of the distribution ($d\ge 8$) 
to a power law $d^{-\mu}$, resulting in $\mu=2.7(1)$.}
        \label{fig:distance_distribiution}
\end{figure}

Next, we present the distribution of flight distances between two
consecutive stops, as shown in Fig.~\ref{fig:distance_distribiution}
for the case of 200 flowers and two bats, here, to fit all bats in,
for a larger system $A=10^6$m$^2$. The histogram is taken
from 100 independent problem instances.
As visible, 
very small distances are rare, because the flower are distributed uniformly
inside the square, therefore small distances do not occur very often,
even if they are overrepresented in optimum schedules.
 The typical flight length is about 2\,m. Large
flight distances are also rare, which is reasonable because of the
optimization task, therefore, if possible, long direct flights are
avoided. Within the double-logarithmic scale, a power law $\sim d^{-\mu}$
is visible
for large distances. From a fit we obtained an exponent $\mu=2.7(1)$.
Note that the underlying distribution of distances, for a uniform random
distributions of the flowers in the plane, is growing $P_0(d)~\sim d$. Therefore,
a power-law distribution which decreases with distance is not evident from
the distribution of underlying distances. Nevertheless,
a power-law (``L\'evy-flight'') behavior is common for experimentally measured
 distribution of distances of natural
foragers \cite{cole1995,viswanathan1996,humphries2010,raichlen2014}.  
Thus, such a power-law  distribution was used in an analytically solvable 
``local-search''  model \cite{viswanathan1999} of harvesting with refilling
resources. Within the model,
a forager either moves to a close by resource if a filled one
is available, and where otherwise the forager performs a long
step with a random direction and a power-law distributed step length.
The value of the exponent was determined from optimizing the
efficiency, which resulted in an exponent $\mu=2$ ,
for the case the exact locations of resources in not known ($\mu<2$ otherwise).
This exponent is smaller than the result obtained here, i.e., the
distribution for NHP decreases faster. 
The reason that very long
distances are suppressed in the NHP with respect to the local search case, 
is that the global optimization target of NHP takes the full environment
into account, leading to shorter solutions and avoiding long distances.

\begin{figure}[!t]
        \centering
\includegraphics[width=0.95\columnwidth]{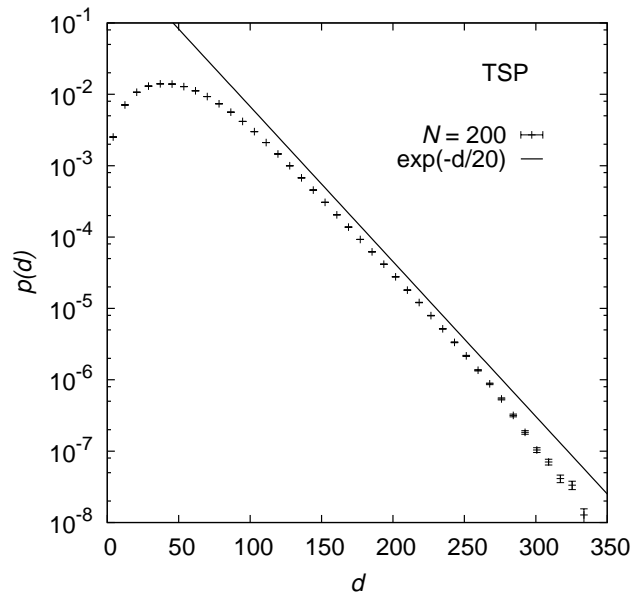}
        \caption{Distribution of travel distances $d$ between two stops
for a TSP solution of a large system with 200 flowers, averaged over
$10^4$ instances. Please note the linear scale for the distances. The line
shows the function $\exp(-d/20)$, as a guide to the eyes.}
        \label{fig:distance_distribiution_TSP}
\end{figure}

For a further comparison, we have also obtained the distribution of distances
for the standard TSP case, here, for $N=200$ flowers distributed in
the same square of side length $10^3$\,m. The result is shown in 
Fig.~\ref{fig:distance_distribiution_TSP}. The behavior in the tail
is very different, an exponential behavior is visible. 
We verified that this is not a finite size effect, by studying a much
larger system with $N=4096$ flowers, where we again found an exponential
tail. The reason is apparently,
due to the constant refilling of the flowers, for the time-dependent NHP
it is more beneficial to take few long flights several times, as compared
to the TSP, where each flower is visited exactly one time and it
is not important when. Anyway, the addition of the biological characteristics
of NHP  like multiple harvesters and time-dependence 
make it apparently fundamentally different from the TSP. Hence, 
in general, studying
ensembles of realistic optimization problems might lead in many cases to 
insights
which cannot be provided solely by studying classical and very abstract
optimization problems.

Also, we studied the temporal patterns of visits at flowers during  
one night for a direct comparison of simulation and experimental data.
 These data are explained in detail below, 
together with the corresponding experimental results,
see Sec.~\ref{sec:temporal}.

\subsection{Phase transition}

\begin{figure}[!t]
        \centering
\includegraphics[width=0.95\columnwidth]{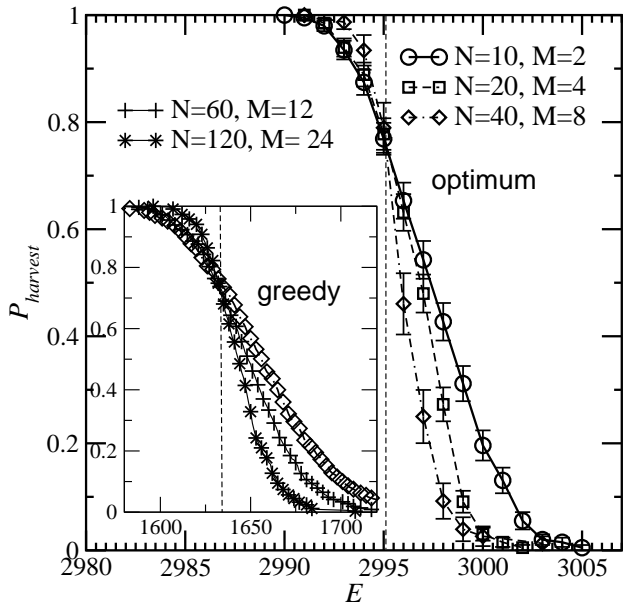}
        \caption{The probability that a system can effectively 
collect a given net energy $E$ per bat, i.e.\ $f(x)\ge E$, 
for different system sizes. $N$ is the number of flowers and $M$ 
the number of bats. 
The inset shows the result for a greedy optimization.
Lines are visual guides only.}
        \label{fig:phasetransition_full}
\end{figure}

For each given  random distribution of the flowers
we obtained the maximum amount $f^{\rm opt} = \max_{x} f(x)$
of the net energy, optimized over all schedules. We measured 
the probability $P_{\rm harvest}$ that
 the bats can effectively 
harvest at least some given amount $E$ of energy per bat, 
i.e.\ the probability that $f^{\rm opt}\ge E$.
 We have performed corresponding simulations for system sizes
10 flowers/2 bats, 20/4 and 40/8, so that the number of bats per
flower was kept constant. In Fig.\ \ref{fig:phasetransition_full}, the
result obtained from averaging over 100 flower distributions is shown.
Note that the amount of nectar produced by all flowers per 12-hour night
amounted to 3024$\mu$l per bat.  
Since the bats have to expend energy for flying, the net gain
will always be less than the nectar energy produced.

The different curves cross near $E_C=2995.1(3)\,\mu$l and become steeper
with growing system size $N$. The ``globally-optimizing'' 
bats will be able most of the time
to collect an amount of nectar smaller or equal to $E_C$, which is very close
to the amount of nectar produced during one night. Clearly, the bats will
never be able to harvest all nectar, since each flower will produce some nectar
after the last visit of a bat. From the behavior of the data it is visible
that for very large systems with many
bats and many flowers, keeping the same bat/flower ratio,
we can expect a threshold behavior resembling a physical phase transition 
\cite{stanley1971}, where suddenly a collective
property of the system changes dramatically.
Similar phenomena have recently
 been observed for artificial random ensembles 
of some classical NP-hard problems like TSP, K-SAT
or the vertex-cover problem
\cite{phase-transitions2005}.
The phase transition observed here appears to be of
 interdisciplinary interest for researchers working
on applied optimization and their relation
to typical computational hardness, because
in contrast to these rather academic
cases like TSP and K-SAT, 
the NHP ensemble is a quite realistic and comprehensive 
model of a real-world problem.

Note that the actual existence of a phase transition does not depend
on how well the bats optimize. 
The simulations based on the greedy algorithm (inset of Fig. \ref{fig:phasetransition_full})  lead to lower individual net energy gain
but again a transition is visible, just occurring
 at much lower energy $E_C\approx 1630(5) \mu$l.

\begin{figure}[!ht]
  \centering \includegraphics[width=0.95\columnwidth]{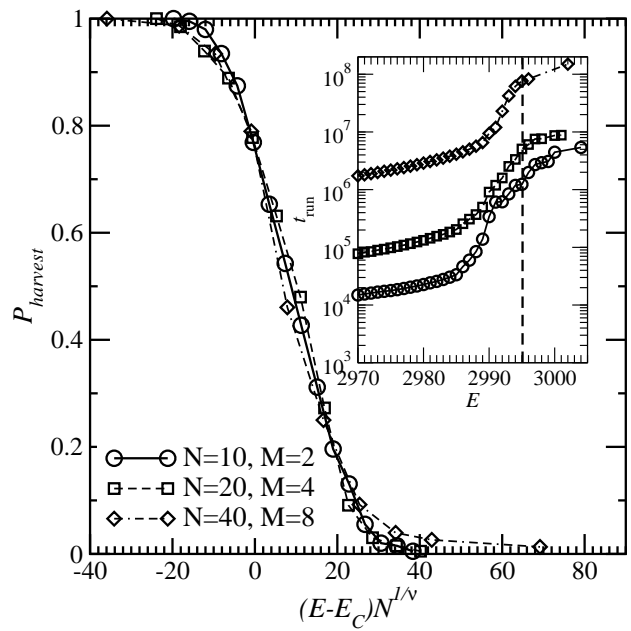}

        \caption{The probability from Fig.\
        \ref{fig:phasetransition_full} plotted for the rescaled energy
        $(E - E_C)\cdot N^{1 / \nu}$ with $\nu = 1.7$.
$N$ is the number of flowers and $M$ 
the number of bats. Lines are visual guides only. The inset
shows the average run time of the algorithm as function of the
        energy for different system sizes. See text for further details.
        \label{fig:scaling_runtime}}
\end{figure}

Next, we characterize this phase transition quantitatively by means of critical
exponents as obtained from {\em finite-size scaling} (FSS)  
\cite{goldenfeld1992,reichl1998}. FSS is a
fundamental approach for overcoming the limited system sizes accessible in 
simulations by studying several ``small'' systems and extrapolating
to large system sizes.
Actually applying FSS works as follows:
First, we determined $E_C$ as the point  where the curves for
different system sizes intersect. Next, we plotted  $P_{\rm harvest}$
 as a function of $\widetilde E=(E-E_C)\cdot N^{1/\nu}$, 
starting with $\nu=1$ and gradually changing the value of $\nu$ such that 
all curves collapse onto one, resulting
in $\nu=1.7(4)$. The result is shown
in Fig.~\ref{fig:scaling_runtime}. 
The fact that the data can be rescaled in this way means that the
phase transition exhibits a growing correlation length,
i.e.\ the optimal tours of the bats depend more and more
on each other when the phase transition point is approached.

The value of the parameter $\nu$ is compatible with the
value $\nu\approx 1.5$ found in simulations of the TSP \cite{gent1996}.
Therefore, with respect to the scaling of the phase transition,
TSP and NHP might be comparable, opposed to the distribution of
distances (see above), where TSP and NHP exhibit non-universality.
Therefore, some characteristics of TSP ``survive'' in the more complicated
multiple-traveler time-dependent NHP.

Next, we turn to the relationship between the phase transition and
the behavior of the optimization algorithm.
 The genetic algorithm used to optimize $f(x)$ gradually
improves the solutions during the optimization.
 Hence at any given so-far running time $t_s$ of the algorithm,
measured in (number of generations)$\times$(size of population),
 there exists a best so-far generated solution 
characterized by the net energy $E$ that is harvested during the corresponding
tours.
In the inset of Fig.\ \ref{fig:scaling_runtime} 
the running time required to gain
a net energy $E$ is shown.
 Interestingly, the running time grows only moderately as a function of $E$
for energy less than $E_C$, but exhibits a drastic increase 
 very close to the phase transition. 
Note also the logarithmically scaled
time axis. Thus, the data indicate that
 the running time increases exponentially
as a function of problem size. Hence, the solvability 
of the NHP seems to be closely related to the typical degree of hardness 
in finding a solution on a computer. This behavior
is similar to the  run time behavior 
recently observed within the statistical-mechanical analysis
of the above mentioned classical optimization problems 
like K-SAT \cite{phase-transitions2005}.


\subsection{Experimental results}

Next, we present the results from the experiments in the rain forest.
Note that these experiments were performed without having the simulation
results already at hand. Therefore, the experiments
 were not designed, e.g., to prove that a phase transition for real bats
exists or that they achieve a harvesting close to optimum (which is an
interesting question for sure, because natural evolution always
aims at the optimum).
Nevertheless, we were able
to observe experimental indications 
 that indeed bats in some way optimize over time and space. Also, we
compare the distribution of activity over time between experiments and
simulations, which yields a remarkable qualitative agreement. 
In the summary, we suggest
refined experimental setups for future expeditions, 
where the simulation results are  taken into account.

\begin{figure}[!h]
        \centering
\includegraphics[width=0.95\columnwidth]{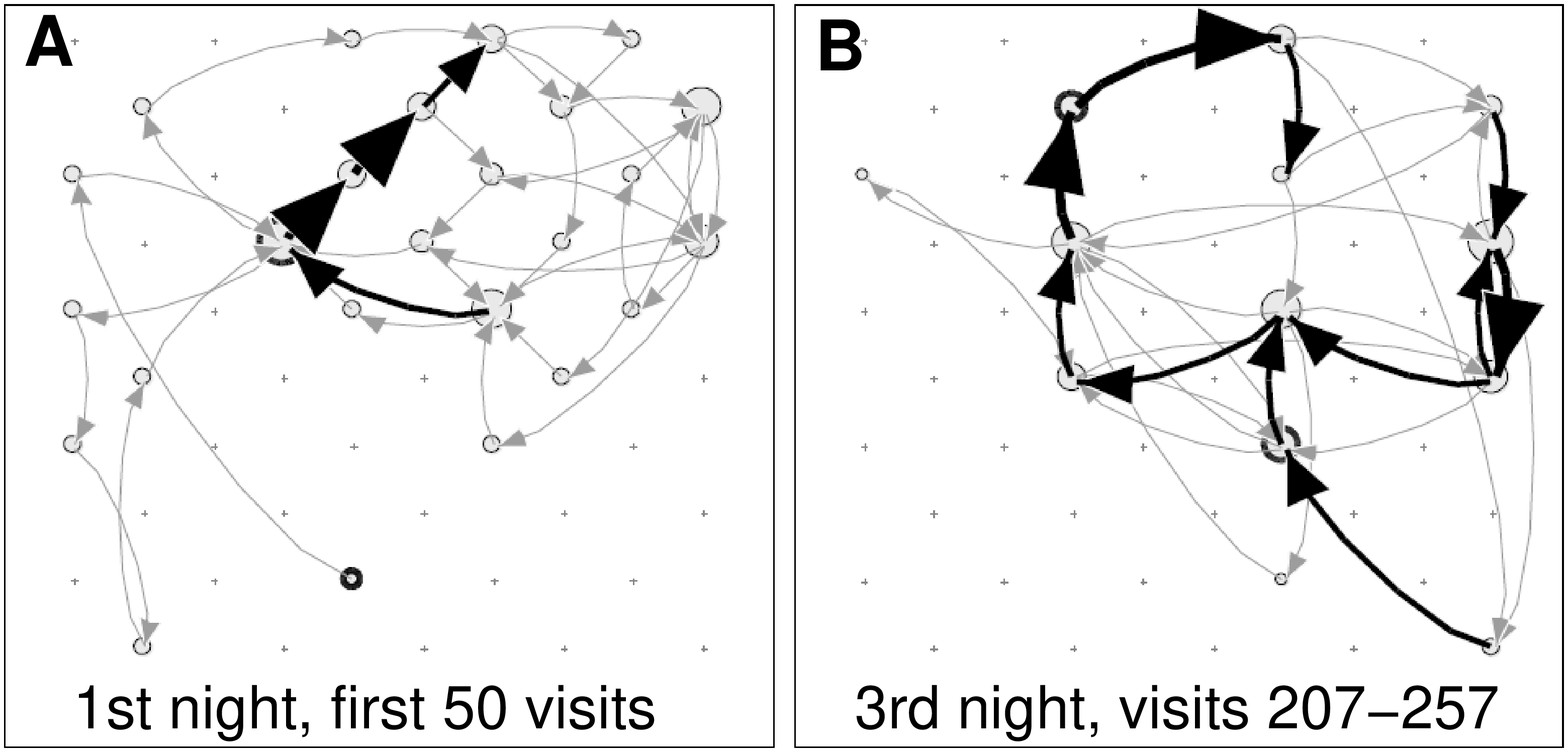}
\includegraphics[width=0.95\columnwidth]{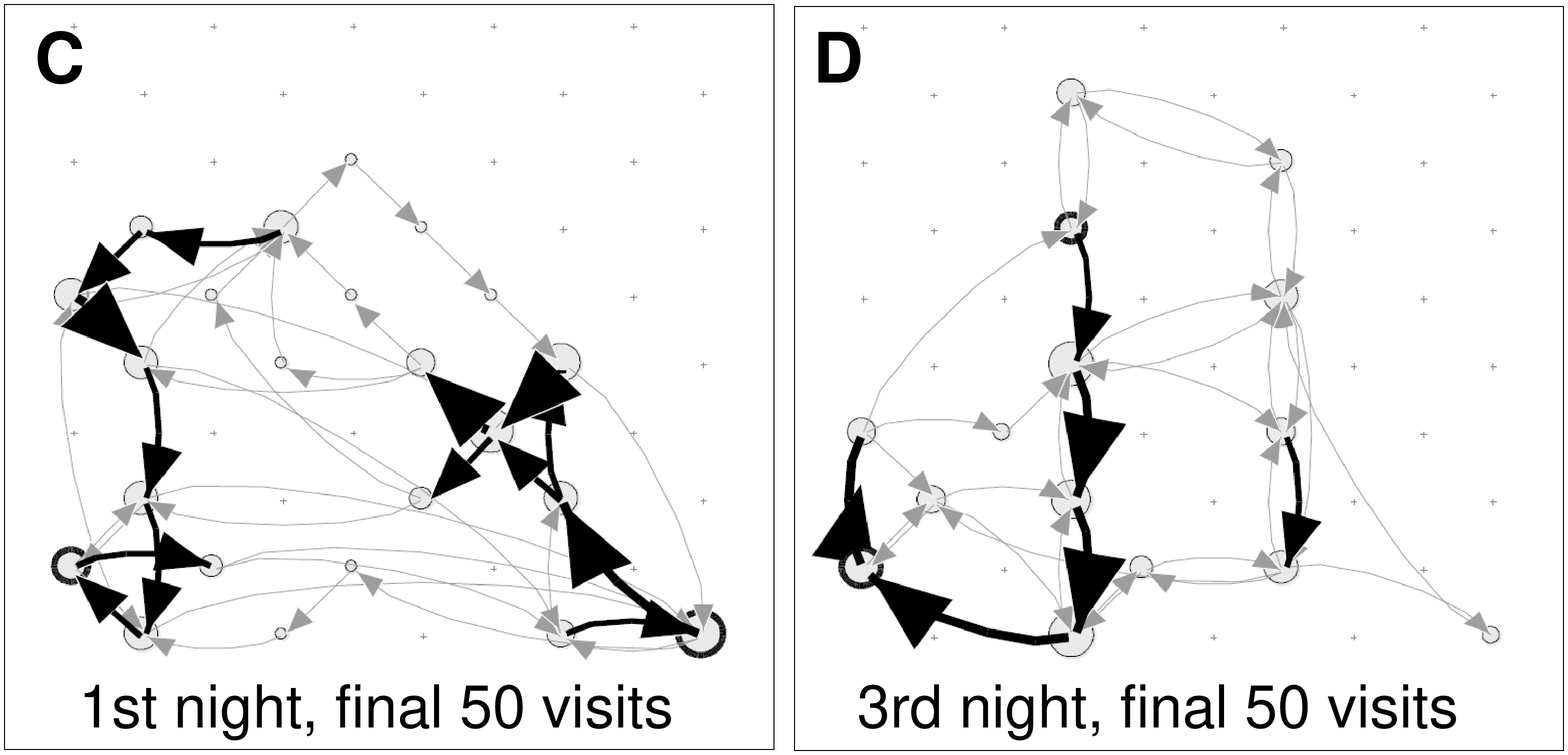}
\vspace*{3mm}
        \caption{The temporal development of flight patterns and space use by two bats (\#141 dark lines, \#178 light lines) 
within three days of visiting the flower field. Flight paths between n = 50 consecutive visits are shown from the beginning (A, C) and the end of three days (B, D) of the experiment. Circle areas are proportional to the number of flower visits, size of arrows are proportional to number of commutes. As visits to QR flowers increased from nights 1 to 3, flight paths changed in direction from diagonal to up-down. 
\label{fig:experimentsR}}
\end{figure}

Here, we analyzed 772 bouts of activity by 8 bats. Feeding bouts lasted
1 to 3 mins during which 7 to 23 flowers were visited and were
followed by intervals of resting for 5 to 25 mins. Bats flew between
experimental flowers for minimum mean distances of 12 to 33 km per
night.  Within our experimental patch of flowers bats did not visit
flowers at random. While initially SR and QR flowers were visited more
or less equally (see Fig.~\ref{fig:experimentsR}A),  
after three days about 80\% of the visits were to
QR flowers (see Fig.~\ref{fig:experimentsR}B).

The spatial pattern of experimental flower distribution allowed us to evaluate 
if bats minimized the lengths of travel paths 
(Fig.\ \ref{fig:experimentsR}). The experimental flowers
 were distributed such that a bat that did not discriminate between flower
 qualities should predominantly travel along diagonal directions to
 minimize travel distance whereas it should travel along up-down directions 
when preferring quickly-refilling flowers. Although we were not able to perform
a large number of experiments to obtain good statistics,
the data indicate that the angle of travel
 directions deviated significantly from random. 
Examples of this development of flight 
pattern are shown in Fig.\ \ref{fig:experimentsR} for 
two bats.
At the onset of 
the experiments the mean angle was apparently closer to a diagonal 
orientation than to an up-down orientation  
(Fig.\ \ref{fig:experimentsR}A, \ref{fig:experimentsR}C)  and 
this angle changed to a predominantly up-down direction towards 
the end of the experiment (Fig.\  \ref{fig:experimentsR}B, 
\ref{fig:experimentsR}D). 
This result shows that the bats 
were both able to distinguish the quality of flowers and minimize 
path lengths (while preferentially sticking to QR flowers) and that they
optimized their behavior with respect to efficiency.

\subsection{\label{sec:temporal}Temporal patterns }

\begin{figure}[!tb]
\begin{center}  
\includegraphics[width=0.95\columnwidth]{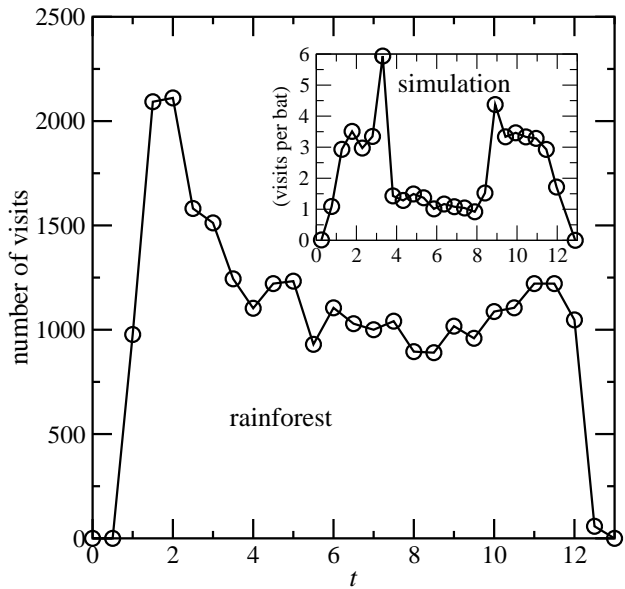}
\end{center}
\caption{
A time distribution (sum of all visits) measured in the rain forest.  
Lines are visual guides only. Inset: Time distribution (visits per bat)
in simulation for instances with $N=40$ resources (flowers)
and $M=2$
bats.  The production rate is triangle-like  and the capacity of each
resource is at $C_k = 200\mu l$. The total  energy produced per flower
and night is $288\mu l$.}
        \label{fig:timedist}
\end{figure}

Finally, we studied the temporal patterns of visits
at flowers during one night. 
The experimental findings as shown in the main plot of Fig.~\ref{fig:timedist}.
  Some bats
remained active in the study area throughout the night whereas other
bats only made shorter visits to the area. Overall, activity was
markedly higher at the beginning of the night and again, with a
smaller peak at the end of the night. 
This can be understood as follows:
At the beginning of the night
a large quantity of nectar has already accumulated from before the onset of
harvesting activity, while towards the end, the bats can again, and with little
effort, harvest non-depleted nectar that has accumulated during the night.
Furthermore, another less important effect is that after an initial harvesting
the stomachs of the bats are full and they need time to digest the nectar.

This can be compared with our numerical results of the optimized
schedules, see inset of Fig.~\ref{fig:timedist}. 
The difference
between active and less-active periods  is more pronounced in the
simulation data: in particular the first peak in the simulation is
shifted towards the middle of the night. This is probably because for
globally optimum harvesting the bats in the simulation need to be
``aware of the future'' such that  they wait until even more nectar
has accumulated and then harvest it from  multiple different resource
locations within short time  intervals. By contrast, real bats at the
end of daily rest are hungry and compete with each other.  Hence,
waiting for too long early at night increases the risk that other bats
deplete the resources. Nevertheless, the qualitative similarity of 
the simulation results with the
data  obtained experimentally in the rain forest shows that our model
makes qualitative predictions of spatial and temporal resource use
that already reproduce the  behavior of real bats.

Note that we also studied numerically the case, where the capacity of
the flowers was set to infinity. Here, to attain the global optimum,
it is worth waiting long until almost the end of the night, to reduce
the total flight distance. And indeed (not shown here), the temporal 
distribution changes to one which grows monotonously and exhibits
one large peak at the end of the night. This shows that the temporal
flight pattern is a direct consequence of optimization performed
by the bats and the limited capacity
of the flowers, which the bats are apparently aware of.


\section{Summary and Discussion}

We have introduced a model to describe the exploitative behavior of
flower visitors harvesting nectar.
 The model is a generalization of the NP-hard 
traveling-salesperson problem, hence no fast algorithm yet exists 
to solve it on a computer. Therefore
 a treatment of the problem using algorithms which
calculate the exact optima is out of reach.
We used genetic algorithms to find optimal or nearly optimal
routes and schedules in numerical simulations 
such that a collection of bats can jointly exploit a habitat globally 
efficiently.

Note that it is not clear a prioi whether 
real bats in a habitat are able to harvest nectar
so efficiently,   primarily because  an ideal-free distribution of
individuals in the spatial and temporal dimension with a complete
avoidance of overlap between all individuals of a bat population is
very unlikely. 
Also, as we observed during our rain forest
studies, a bat’s flight path during a night does not only connect
harvesting events but detours to interact with other bats, search for
new flowers or avoid danger. This makes a direct comparison of
efficiency impossible.  
Nevertheless, as mentioned, collective individual
optimization may lead to global optima, which is exploited by
swarm intelligence algorithms like ant-colony optimization. Therefore,
the study of global optima of the NHP is justified.

In fact, from the preliminary results of our experiments in the rain
forest, we can  conclude that the bats are indeed able to partially
optimize. The experiments are based
on the surveillance of bats using artificial flowers and measuring their
level of activity during the night. The results indicate that
the underlying model exhibits enough ingredients to  sufficiently describe the 
basic behavior of the bats, in particular that
the bats exhibit spatio-temporal memory:
 The bats seem to be able to optimize
their exploitation, which is apparent in the increasing fraction of flights
to ``quickly refilling'' flowers.

In an analysis of the flight lengths of the optimum schedule, we found
a power-law (L\'evy) distribution. Such distributions have often
been observed in various ecological systems, like flies \cite{cole1995},
Albatrosses \cite{viswanathan1996}, marine predators \cite{humphries2010},
or humans \cite{raichlen2014}.
The quest for the 
origin of such behavior arose a great deal of attention and led
to many studies trying to explain its origins \cite{bartumeus2007}.
In modelling harvesting or food search, power-law distributions have been
used severals times, e.g., for determining the optimum exponent
of such a distribution, when the ressource distribution might be unkown
\cite{viswanathan1999}. 
There are other more recent models, where 
a power-law distribution of the step sizes is also assumed in some
way or the other \cite{reynolds2009,gautestad2013,zhao2015,kusmierz2017}.
In all these studies the power-law behavior does not emerge without 
assuming it, but they are explicit part of the models.
 Therefore, such models correspond to the assumption that evolution
has led the animals to develop a (random) 
search strategy which encodes such L\'evy
walks, in particular with the aim to \emph{detect} yet unknown food
ressources.

It is not always true that L\'evy flights lead
to the optimal foraging behavior. Clearly, if the ressources are dense,
ranom walk behavior is suffiecient to find them, while minimizing the
step length. Also,
under certain conditions other distribuions of the step sizes are optimal, 
e.g., a ballistic search \cite{james2008}.

On the other hand, 
a power-law distribution of distances was also found for a foraging
model \cite{maya2017}, where no assumption about the distribution
of step sizes was made, but where the ressources were distributed in the plane
with ressource qualitiy values drawn from a scale-free power-law distribution.
One forgager with a local search strategy was simulated, such that
always the closest ressource was taken with the smallest distance to quality
ration. A power-law distribution for the distance was 
only found with a proper choice of the power for the quality distribution.
Hence, the observed distance distribution seems to be  here a result 
of the specific power-law choice of the
ressource quality  distribution, but not a true emergent property
of the system.

Thus, we find in the present study
a power-law distribution of travel distances without any correspoding
assumption and with the
ressources being distributed uniform randomly. 
In no part of the present model a power-law distribution
is assumed, neither for step-sizes of animal movement nor for
the quantities describing the food ressources. This means also that 
no assumptions about evolutionary processes favoring such 
power-law distributions have been made. Thus, emerging L\'evy-walk
distributions is possible without the assumption of directly
influence of evolution, in contrast to some previous ideas 
\cite{bartumeus2007}. On the other hand, there exist models of
target-searching agents in randomly changing
environments, where an evolution-caused power-law 
distribution was observed \cite{van_dartel2009,wosniack2017}.

Nevertheless, for our model, in principal,
 the animals have perfect knowledge of the food ressources, partially
justified by the good spatio-temporal memory of the bats, we observed
in experiments. Thus the
observed L\'evy distribution is not describing a random search strategy, but it
emerges without specific assumptions just from optimizing the search
efforts versus search gain. Interestingly, the time dependence, 
seems to be an important ingredient, because for the TSP we observe
an exponential distribution of step sizes for the optimal tours.

Furthermore, beyond the statistical properties of the optimal schedules,
when increasing the demand of the bats in our simulations, 
we found a phase transition from a region where
the bats have enough nectar available to a region where not all bats
can be fed sufficiently. This, as mentioned above,
is due to a growing correlation length of the system.
 From a biological point of view this means that optimization becomes more 
demanding as more individuals are in exploitation competition. This is 
because individual optimization becomes increasingly dependent on how well an individual can avoid the effects of exploitative competition caused by the 
other individuals. This we could observe also when studying the
running time of the genetic algorithm:
The transition coincides with a marked increase in 
running time, indicating that the problem
becomes typically hard to solve on a computer.

Note that  we have  observed the phase transition, coinciding
with a substantial increase in running time, for 
an ensemble of realistic time-dependent combinatorial optimization
problems, in contrast to previously studied models which are rather
simple and static, like random K-Satisfiability
or the coloring problem on random graphs. 
This indicates that also for realistic optimization problems
phase transitions occur 
and it may be useful to investigate them, e.g., to
determine optimum working conditions by analyzing
ratios of outcome versus effort, which are often
found to be best near phase transitions.
In particular, as visible in the distribution of flight distances,
the results indicate that the behavior of NHP, which involves
multiple bats and time-dependence, is different from the strongly
reduced TSP model. Nevertheless, the growth of the correlation,
as measured by the critical exponent $\nu$, appears to be similar.

Concerning future research directions,
the occurrence of a phase transition in the optimum solution
could serve as a guiding line for also further experimental studies:
 It is tempting to ask, whether such a phase transition can be
observed experimentally in the natural rain forest environment. 
Within an experiment using artificial flowers,
for example, the amount of nectar offered until the bats
start to move out of the habitat could be continuously reduced.
Note that with the present experimental data, where the feeders
were controlled such that the bats acted independently of each other,
it is unfortunately not a simple matter of re\-ana\-lysis of the data. There,
new experimental (and, unfortunately, involved and expensive)
campaigns  in the rain forest, guided by the present result
and forcing the bats to compete,  will be needed.
 If a similar phase transition is observed
experimentally, the  value $E_C^{\rm exp}$ where the phase transition 
occurs would indicate
how well the bats are able to optimize, compared with the theoretical optimum
studied here. In particular, if the experimental data for different ``sizes'',
i.e., number of bats and flowers, can be rescaled (as in the simulation
results in Fig.\ \ref{fig:scaling_runtime}) this would mean
that there is a growing correlation between the bats. This would provide
strong evidence that the bats communicate or interact in some
way, directly or maybe indirectly through flower exploitation,
 leading to a jointly more efficient exploitation of the habitat.
Also, the analysis of the schedules of competing bats, will show whether
they partition the habitat among them, as seen in the optimal schedules,
and whether such a partition evolves over time.
In general, performing such experiments having a comparison with the optimum
will result in a much greater
 understanding of the
exploitation capabilities of  highly-evolved vertebrates.

On the simulation side, one could perform extensive 
detailed agent-based simulations,
 with bats having local memory and (possibly different) 
local optimization strategies.
This is motivated by other results from our experiments (not shown),
where we grouped the bat's activities into bouts and measured the
frequency $f$ bats revisited flowers within a bout, as a function of
the number $n_{\rm vis}$ 
of visits in a bout. We found that, the more visits are included
in a bout, the more likely revisits occur, i.e. $f(n_{\rm vis})$
exhibits a sigmoid shape. This inidcate that bats exhibit a limited
working memory helping them to harvest efficiently.
We performed some preliminary 
agent based simulations with bats performing greedy search for the 
next flower to visit. The bats had
a limited amount of working memory allowing them to avoid flowers they
visited recently. We found that for bats remembering two or three past visits,
the $f(n_{\rm vis})$ function resembled the measured data best.
This result makes it tempting to perform agent based simulations
more extensively.
Given that we found a phase transition for the optimum as well as for
the very simple greedy solutions, it appears likely that it exists
for the agent-based case as well. It should be of high interest to know
if and how the transition depends on the ``cognitive complexity'' 
of the agents. By carefully
adjusting the experiments to the agent's complexity, one might even
try to determine how well-developed the ``algorithms'' for spatial
and temporal cognition in natural bats are.

\section{Acknowledgments}
MJ, AKH and YW were supported 
 by the {\em VolkswagenStiftung} (Germany) within the program
``Nachwuchsgruppen an Universit\"aten''.
HS was supported by the German Science Foundation (DFG) through
        the grant HA 3169/8-1.
YW is supported by the DFG Cluster of Excellence ``Neurocure''.
The simulations were partially performed at computing clusters at the
``Gesellschaft f\"ur Wissenschaftliche Datenverarbeitung'' and at
the ``Institute of Theoretical Physics'' in G\"ottingen.
The simulations were also partially performed at the HPC cluster CARL, 
        located at the University of Oldenburg (Germany) and funded by the DFG
        through its Major Research Instrumentation Programme
        (INST 184/157-1 FUGG) and the Ministry of
        Science and Culture (MWK) of the Lower Saxony State.

\bibliography{alex_refs}

\begin{thebibliography}{41}
\expandafter\ifx\csname natexlab\endcsname\relax\def\natexlab#1{#1}\fi
\expandafter\ifx\csname bibnamefont\endcsname\relax
  \def\bibnamefont#1{#1}\fi
\expandafter\ifx\csname bibfnamefont\endcsname\relax
  \def\bibfnamefont#1{#1}\fi
\expandafter\ifx\csname citenamefont\endcsname\relax
  \def\citenamefont#1{#1}\fi
\expandafter\ifx\csname url\endcsname\relax
  \def\url#1{\texttt{#1}}\fi
\expandafter\ifx\csname urlprefix\endcsname\relax\def\urlprefix{URL }\fi
\providecommand{\bibinfo}[2]{#2}
\providecommand{\eprint}[2][]{\url{#2}}

\bibitem[{\citenamefont{Stephens et~al.}(2007)\citenamefont{Stephens, Brown,
  and Ydenberg}}]{stephens2007}
\bibinfo{author}{\bibfnamefont{D.~W.} \bibnamefont{Stephens}},
  \bibinfo{author}{\bibfnamefont{J.~S.} \bibnamefont{Brown}}, \bibnamefont{and}
  \bibinfo{author}{\bibfnamefont{R.~C.} \bibnamefont{Ydenberg}},
  \emph{\bibinfo{title}{Foraging: Behavior and Ecology}}
  (\bibinfo{publisher}{University of Chicago Press},
  \bibinfo{address}{Chicago}, \bibinfo{year}{2007}).

\bibitem[{\citenamefont{Prins and van Langevelde}(2008)}]{prins2008}
\bibinfo{author}{\bibfnamefont{H.~H.~T.} \bibnamefont{Prins}} \bibnamefont{and}
  \bibinfo{author}{\bibfnamefont{F.}~\bibnamefont{van Langevelde}},
  \emph{\bibinfo{title}{Resource Ecology: Spatial and Temporal Dynamics of
  Foraging}} (\bibinfo{publisher}{Springer}, \bibinfo{address}{Heidelberg},
  \bibinfo{year}{2008}).

\bibitem[{\citenamefont{Liu and Passino}(2002)}]{liu2002}
\bibinfo{author}{\bibfnamefont{Y.}~\bibnamefont{Liu}} \bibnamefont{and}
  \bibinfo{author}{\bibfnamefont{K.}~\bibnamefont{Passino}},
  \bibinfo{journal}{J. Optim. Theory Appl.} \textbf{\bibinfo{volume}{115}},
  \bibinfo{pages}{603} (\bibinfo{year}{2002}).

\bibitem[{\citenamefont{Viswanathan et~al.}(1996)\citenamefont{Viswanathan,
  Afanasyev, Buldyrev, Murphy, Prince, and Stanley}}]{viswanathan1996}
\bibinfo{author}{\bibfnamefont{G.~M.} \bibnamefont{Viswanathan}},
  \bibinfo{author}{\bibfnamefont{V.}~\bibnamefont{Afanasyev}},
  \bibinfo{author}{\bibfnamefont{S.~V.} \bibnamefont{Buldyrev}},
  \bibinfo{author}{\bibfnamefont{E.~J.} \bibnamefont{Murphy}},
  \bibinfo{author}{\bibfnamefont{P.~A.} \bibnamefont{Prince}},
  \bibnamefont{and} \bibinfo{author}{\bibfnamefont{H.~E.}
  \bibnamefont{Stanley}}, \bibinfo{journal}{Nature}
  \textbf{\bibinfo{volume}{381}}, \bibinfo{pages}{413} (\bibinfo{year}{1996}).

\bibitem[{\citenamefont{van Dartel et~al.}(2004)\citenamefont{van Dartel,
  Postma, van~den Herik, and de~Croon}}]{van_dartel2009}
\bibinfo{author}{\bibfnamefont{M.}~\bibnamefont{van Dartel}},
  \bibinfo{author}{\bibfnamefont{E.}~\bibnamefont{Postma}},
  \bibinfo{author}{\bibfnamefont{J.}~\bibnamefont{van~den Herik}},
  \bibnamefont{and} \bibinfo{author}{\bibfnamefont{G.}~\bibnamefont{de~Croon}},
  \bibinfo{journal}{Connection Science} \textbf{\bibinfo{volume}{16}},
  \bibinfo{pages}{169} (\bibinfo{year}{2004}).

\bibitem[{\citenamefont{Wosniack et~al.}(2017)\citenamefont{Wosniack, Santos,
  Raposo, Viswanathan, and da~Luz}}]{wosniack2017}
\bibinfo{author}{\bibfnamefont{M.~E.} \bibnamefont{Wosniack}},
  \bibinfo{author}{\bibfnamefont{M.~C.} \bibnamefont{Santos}},
  \bibinfo{author}{\bibfnamefont{E.~P.} \bibnamefont{Raposo}},
  \bibinfo{author}{\bibfnamefont{G.~M.} \bibnamefont{Viswanathan}},
  \bibnamefont{and} \bibinfo{author}{\bibfnamefont{M.~G.~E.}
  \bibnamefont{da~Luz}}, \bibinfo{journal}{PLOS Computational Biology}
  \textbf{\bibinfo{volume}{13}}, \bibinfo{pages}{1} (\bibinfo{year}{2017}).

\bibitem[{\citenamefont{Hartmann and Weigt}(2005)}]{phase-transitions2005}
\bibinfo{author}{\bibfnamefont{A.~K.} \bibnamefont{Hartmann}} \bibnamefont{and}
  \bibinfo{author}{\bibfnamefont{M.}~\bibnamefont{Weigt}},
  \emph{\bibinfo{title}{Phase Transitions in Combinatorial Optimization
  Problems}} (\bibinfo{publisher}{Wiley-VCH}, \bibinfo{address}{Weinheim},
  \bibinfo{year}{2005}).

\bibitem[{\citenamefont{Mitchell et~al.}(1992)\citenamefont{Mitchell, Selman,
  and Levesque}}]{mitchell1992}
\bibinfo{author}{\bibfnamefont{D.~G.} \bibnamefont{Mitchell}},
  \bibinfo{author}{\bibfnamefont{B.}~\bibnamefont{Selman}}, \bibnamefont{and}
  \bibinfo{author}{\bibfnamefont{H.}~\bibnamefont{Levesque}}, in
  \emph{\bibinfo{booktitle}{Proc. 10th Natl. Conf. Artif. Intell. (AAAI-92)}}
  (\bibinfo{publisher}{AAAI Press/MIT Press}, \bibinfo{address}{Cambridge},
  \bibinfo{year}{1992}), pp. \bibinfo{pages}{440--446}.

\bibitem[{\citenamefont{Gent and Walsh}(1996)}]{gent1996}
\bibinfo{author}{\bibfnamefont{I.~P.} \bibnamefont{Gent}} \bibnamefont{and}
  \bibinfo{author}{\bibfnamefont{T.}~\bibnamefont{Walsh}}, in
  \emph{\bibinfo{booktitle}{Proceedings of 12th European Conference on
  Artificial Intelligence. ECAI '96}}, edited by
  \bibinfo{editor}{\bibfnamefont{W.}~\bibnamefont{Wahlster}}
  (\bibinfo{publisher}{Wiley}, \bibinfo{address}{Chichester},
  \bibinfo{year}{1996}), p. \bibinfo{pages}{170}.

\bibitem[{\citenamefont{Moore and Mertens}(2011)}]{moore2011}
\bibinfo{author}{\bibfnamefont{C.}~\bibnamefont{Moore}} \bibnamefont{and}
  \bibinfo{author}{\bibfnamefont{S.}~\bibnamefont{Mertens}},
  \emph{\bibinfo{title}{The Nature of Computation}} (\bibinfo{publisher}{Oxford
  University Press}, \bibinfo{address}{Oxford}, \bibinfo{year}{2011}).

\bibitem[{\citenamefont{Papadimitriou and Steiglitz}(2000)}]{papadimitriou2000}
\bibinfo{author}{\bibfnamefont{C.}~\bibnamefont{Papadimitriou}}
  \bibnamefont{and}
  \bibinfo{author}{\bibfnamefont{K.}~\bibnamefont{Steiglitz}},
  \emph{\bibinfo{title}{Combinatorial Optimization -- Algorithms and
  Complexity}} (\bibinfo{publisher}{Dover Publications Inc.},
  \bibinfo{address}{Mineola, NY}, \bibinfo{year}{2000}).

\bibitem[{\citenamefont{Gutin}(2002)}]{gutin2002}
\bibinfo{author}{\bibfnamefont{G.}~\bibnamefont{Gutin}},
  \emph{\bibinfo{title}{The Traveling Salesman Problem and Its Variations}}
  (\bibinfo{publisher}{Kluwer Academic Publishers},
  \bibinfo{address}{Dordrecht}, \bibinfo{year}{2002}).

\bibitem[{\citenamefont{Garey and Johnson}(1979)}]{garey1979}
\bibinfo{author}{\bibfnamefont{M.~R.} \bibnamefont{Garey}} \bibnamefont{and}
  \bibinfo{author}{\bibfnamefont{D.~S.} \bibnamefont{Johnson}},
  \emph{\bibinfo{title}{Computers and intractability}}
  (\bibinfo{publisher}{W.H. Freemann}, \bibinfo{address}{San Francisco},
  \bibinfo{year}{1979}).

\bibitem[{\citenamefont{Dorigo and Stützle}(2004)}]{dorigo2004}
\bibinfo{author}{\bibfnamefont{M.}~\bibnamefont{Dorigo}} \bibnamefont{and}
  \bibinfo{author}{\bibfnamefont{T.}~\bibnamefont{Stützle}},
  \emph{\bibinfo{title}{Ant Colony Optimization}} (\bibinfo{publisher}{MIT
  Press}, \bibinfo{address}{Cambridge (USA)}, \bibinfo{year}{2004}).

\bibitem[{\citenamefont{Pintea}(2014)}]{pintea2014}
\bibinfo{author}{\bibfnamefont{C.-M.} \bibnamefont{Pintea}},
  \emph{\bibinfo{title}{Advances in Bio-inspired Computing for Combinatorial
  Optimization Problem}} (\bibinfo{publisher}{Springer},
  \bibinfo{address}{Heidelberg}, \bibinfo{year}{2014}).

\bibitem[{\citenamefont{Winter and von Helversen}(2001)}]{winter2001}
\bibinfo{author}{\bibfnamefont{Y.}~\bibnamefont{Winter}} \bibnamefont{and}
  \bibinfo{author}{\bibfnamefont{O.}~\bibnamefont{von Helversen}}, in
  \emph{\bibinfo{booktitle}{Cognitive Ecology of Pollination}}
  (\bibinfo{publisher}{Cambridge University Press},
  \bibinfo{address}{Cambridge}, \bibinfo{year}{2001}), pp.
  \bibinfo{pages}{148--170}.

\bibitem[{\citenamefont{Helversen and Winter}(2003)}]{helversen2003}
\bibinfo{author}{\bibfnamefont{O.}~\bibnamefont{Helversen}} \bibnamefont{and}
  \bibinfo{author}{\bibfnamefont{Y.}~\bibnamefont{Winter}}, in
  \emph{\bibinfo{booktitle}{Bat ecology}}, edited by
  \bibinfo{editor}{\bibfnamefont{T.~H.} \bibnamefont{Kunz}} \bibnamefont{and}
  \bibinfo{editor}{\bibfnamefont{B.~M.} \bibnamefont{Fenton}}
  (\bibinfo{publisher}{University of Chicago Press},
  \bibinfo{address}{Chicago}, \bibinfo{year}{2003}), pp.
  \bibinfo{pages}{346--397}.

\bibitem[{\citenamefont{Cole}(1995)}]{cole1995}
\bibinfo{author}{\bibfnamefont{B.~J.} \bibnamefont{Cole}},
  \bibinfo{journal}{Animal Behaviour} \textbf{\bibinfo{volume}{50}},
  \bibinfo{pages}{1317 } (\bibinfo{year}{1995}), ISSN
  \bibinfo{issn}{0003-3472}.

\bibitem[{\citenamefont{Humphries et~al.}(2010)\citenamefont{Humphries,
  Queiroz, Dyer, Pade, Musyl, Fuller, Brunnschweiler, Doyle, Houghton, Hays
  et~al.}}]{humphries2010}
\bibinfo{author}{\bibfnamefont{N.~E.} \bibnamefont{Humphries}},
  \bibinfo{author}{\bibfnamefont{N.}~\bibnamefont{Queiroz}},
  \bibinfo{author}{\bibfnamefont{J.~R.~M.} \bibnamefont{Dyer}},
  \bibinfo{author}{\bibfnamefont{N.~G.} \bibnamefont{Pade}},
  \bibinfo{author}{\bibfnamefont{K.~M.} \bibnamefont{Musyl},
  \bibfnamefont{Michael K.~Schaefer}}, \bibinfo{author}{\bibfnamefont{D.~W.}
  \bibnamefont{Fuller}}, \bibinfo{author}{\bibfnamefont{J.~M.}
  \bibnamefont{Brunnschweiler}}, \bibinfo{author}{\bibfnamefont{T.~K.}
  \bibnamefont{Doyle}}, \bibinfo{author}{\bibfnamefont{J.~D.~R.}
  \bibnamefont{Houghton}}, \bibinfo{author}{\bibfnamefont{G.~C.}
  \bibnamefont{Hays}}, \bibnamefont{et~al.}, \bibinfo{journal}{Nature}
  \textbf{\bibinfo{volume}{465}}, \bibinfo{pages}{1066} (\bibinfo{year}{2010}).

\bibitem[{\citenamefont{Raichlen et~al.}(2014)\citenamefont{Raichlen, Wood,
  Gordon, Mabulla, Marlowe, and Pontzer}}]{raichlen2014}
\bibinfo{author}{\bibfnamefont{D.~A.} \bibnamefont{Raichlen}},
  \bibinfo{author}{\bibfnamefont{B.~M.} \bibnamefont{Wood}},
  \bibinfo{author}{\bibfnamefont{A.~D.} \bibnamefont{Gordon}},
  \bibinfo{author}{\bibfnamefont{A.~Z.~P.} \bibnamefont{Mabulla}},
  \bibinfo{author}{\bibfnamefont{F.~W.} \bibnamefont{Marlowe}},
  \bibnamefont{and} \bibinfo{author}{\bibfnamefont{H.}~\bibnamefont{Pontzer}},
  \bibinfo{journal}{PNAS} \textbf{\bibinfo{volume}{111}}, \bibinfo{pages}{728}
  (\bibinfo{year}{2014}).

\bibitem[{\citenamefont{Viswanathan et~al.}(1999)\citenamefont{Viswanathan,
  Buldyrev, Havlin, da~Lu, Raposo, and Stanley}}]{viswanathan1999}
\bibinfo{author}{\bibfnamefont{G.~M.} \bibnamefont{Viswanathan}},
  \bibinfo{author}{\bibfnamefont{S.~V.} \bibnamefont{Buldyrev}},
  \bibinfo{author}{\bibfnamefont{S.}~\bibnamefont{Havlin}},
  \bibinfo{author}{\bibfnamefont{M.~G.~E.} \bibnamefont{da~Lu}},
  \bibinfo{author}{\bibfnamefont{E.~P.} \bibnamefont{Raposo}},
  \bibnamefont{and} \bibinfo{author}{\bibfnamefont{H.~E.}
  \bibnamefont{Stanley}}, \bibinfo{journal}{Nature}
  \textbf{\bibinfo{volume}{401}}, \bibinfo{pages}{911} (\bibinfo{year}{1999}).

\bibitem[{\citenamefont{Reynolds and Bartumeus}(2009)}]{reynolds2009}
\bibinfo{author}{\bibfnamefont{A.~M.} \bibnamefont{Reynolds}} \bibnamefont{and}
  \bibinfo{author}{\bibfnamefont{F.}~\bibnamefont{Bartumeus}},
  \bibinfo{journal}{J. Theor. Biol.} \textbf{\bibinfo{volume}{260}},
  \bibinfo{pages}{98 } (\bibinfo{year}{2009}), ISSN \bibinfo{issn}{0022-5193}.

\bibitem[{\citenamefont{Gautestad and Mysterud}(2013)}]{gautestad2013}
\bibinfo{author}{\bibfnamefont{A.~O.} \bibnamefont{Gautestad}}
  \bibnamefont{and} \bibinfo{author}{\bibfnamefont{A.}~\bibnamefont{Mysterud}},
  \bibinfo{journal}{Movement Ecology} \textbf{\bibinfo{volume}{1}},
  \bibinfo{pages}{9} (\bibinfo{year}{2013}).

\bibitem[{\citenamefont{Zhao et~al.}(2015)\citenamefont{Zhao, Jurdak, Liu,
  Westcott, Kusy, Parry, Sommer, and McKeown}}]{zhao2015}
\bibinfo{author}{\bibfnamefont{K.}~\bibnamefont{Zhao}},
  \bibinfo{author}{\bibfnamefont{R.}~\bibnamefont{Jurdak}},
  \bibinfo{author}{\bibfnamefont{J.}~\bibnamefont{Liu}},
  \bibinfo{author}{\bibfnamefont{D.}~\bibnamefont{Westcott}},
  \bibinfo{author}{\bibfnamefont{B.}~\bibnamefont{Kusy}},
  \bibinfo{author}{\bibfnamefont{H.}~\bibnamefont{Parry}},
  \bibinfo{author}{\bibfnamefont{P.}~\bibnamefont{Sommer}}, \bibnamefont{and}
  \bibinfo{author}{\bibfnamefont{A.}~\bibnamefont{McKeown}},
  \bibinfo{journal}{J. Royal Soc. Interf.} \textbf{\bibinfo{volume}{12}}
  (\bibinfo{year}{2015}), \eprint{20141158}.

\bibitem[{\citenamefont{{Ku{\'s}mierz} and {Toyoizumi}}(2017)}]{kusmierz2017}
\bibinfo{author}{\bibfnamefont{{\L}.}~\bibnamefont{{Ku{\'s}mierz}}}
  \bibnamefont{and}
  \bibinfo{author}{\bibfnamefont{T.}~\bibnamefont{{Toyoizumi}}},
  \bibinfo{journal}{ArXiv e-prints}  (\bibinfo{year}{2017}),
  \eprint{1710.01889}.

\bibitem[{\citenamefont{F.}(2007)}]{bartumeus2007}
\bibinfo{author}{\bibfnamefont{B.}~\bibnamefont{F.}},
  \bibinfo{journal}{Fractals} \textbf{\bibinfo{volume}{15}},
  \bibinfo{pages}{151} (\bibinfo{year}{2007}).

\bibitem[{\citenamefont{Winter}(1999)}]{winter1999}
\bibinfo{author}{\bibfnamefont{Y.}~\bibnamefont{Winter}}, \bibinfo{journal}{J.
  Experim. Biol.} \textbf{\bibinfo{volume}{202}}, \bibinfo{pages}{1917}
  (\bibinfo{year}{1999}).

\bibitem[{\citenamefont{Thiele and Winter}(2005)}]{thiele2005}
\bibinfo{author}{\bibfnamefont{J.}~\bibnamefont{Thiele}} \bibnamefont{and}
  \bibinfo{author}{\bibfnamefont{Y.}~\bibnamefont{Winter}},
  \bibinfo{journal}{Animal Behaviour} \textbf{\bibinfo{volume}{69}},
  \bibinfo{pages}{315} (\bibinfo{year}{2005}).

\bibitem[{\citenamefont{Winter and Stich}(2005)}]{winter2005}
\bibinfo{author}{\bibfnamefont{Y.}~\bibnamefont{Winter}} \bibnamefont{and}
  \bibinfo{author}{\bibfnamefont{K.~P.} \bibnamefont{Stich}},
  \bibinfo{journal}{J. Experim. Biol.} \textbf{\bibinfo{volume}{208}},
  \bibinfo{pages}{539} (\bibinfo{year}{2005}).

\bibitem[{\citenamefont{Hartmann}(2015)}]{practical_guide2015}
\bibinfo{author}{\bibfnamefont{A.~K.} \bibnamefont{Hartmann}},
  \emph{\bibinfo{title}{{Big Practical Guide to Computer Simulations}}}
  (\bibinfo{publisher}{World Scientific}, \bibinfo{address}{Singapore},
  \bibinfo{year}{2015}).

\bibitem[{\citenamefont{Goldberg}(1989)}]{goldberg1989}
\bibinfo{author}{\bibfnamefont{D.~E.} \bibnamefont{Goldberg}},
  \emph{\bibinfo{title}{Genetic Algorithms in Search, Optimization and Machine
  Learning}} (\bibinfo{publisher}{Addison-Wesley}, \bibinfo{address}{Reading
  (MA)}, \bibinfo{year}{1989}).

\bibitem[{\citenamefont{Goldberg and Deb}(1991)}]{goldberg1991}
\bibinfo{author}{\bibfnamefont{D.~E.} \bibnamefont{Goldberg}} \bibnamefont{and}
  \bibinfo{author}{\bibfnamefont{K.}~\bibnamefont{Deb}}, in
  \emph{\bibinfo{booktitle}{Foundations of Genetic Algorithms}}, edited by
  \bibinfo{editor}{\bibfnamefont{G.~J.~E.} \bibnamefont{Rawlings}}
  (\bibinfo{publisher}{Addison-Wesley}, \bibinfo{address}{Reading (MA)},
  \bibinfo{year}{1991}), pp. \bibinfo{pages}{69--93}.

\bibitem[{\citenamefont{Blickle and Thiele}(1995)}]{blickle1995}
\bibinfo{author}{\bibfnamefont{T.}~\bibnamefont{Blickle}} \bibnamefont{and}
  \bibinfo{author}{\bibfnamefont{L.}~\bibnamefont{Thiele}},
  \bibinfo{type}{TIK-Report}, \bibinfo{institution}{ETH Z\"urich}
  (\bibinfo{year}{1995}).

\bibitem[{\citenamefont{Bektas}(2003)}]{bektas2003}
\bibinfo{author}{\bibfnamefont{T.}~\bibnamefont{Bektas}},
  \bibinfo{journal}{Omega} \textbf{\bibinfo{volume}{34}}, \bibinfo{pages}{209}
  (\bibinfo{year}{2003}).

\bibitem[{\citenamefont{Applegate et~al.}(2003)\citenamefont{Applegate, Bixby,
  Chv{\'a}tal, and Cook}}]{applegate2003}
\bibinfo{author}{\bibfnamefont{D.}~\bibnamefont{Applegate}},
  \bibinfo{author}{\bibfnamefont{R.}~\bibnamefont{Bixby}},
  \bibinfo{author}{\bibfnamefont{V.}~\bibnamefont{Chv{\'a}tal}},
  \bibnamefont{and} \bibinfo{author}{\bibfnamefont{W.}~\bibnamefont{Cook}},
  \bibinfo{journal}{Math. Program.} \textbf{\bibinfo{volume}{97}},
  \bibinfo{pages}{91} (\bibinfo{year}{2003}).

\bibitem[{\citenamefont{Thiele}(2006)}]{thiele2006}
\bibinfo{author}{\bibfnamefont{J.}~\bibnamefont{Thiele}},
  \bibinfo{type}{dissertation}, \bibinfo{school}{Ludwig Maximilians
  University}, \bibinfo{address}{Munich} (\bibinfo{year}{2006}).

\bibitem[{\citenamefont{Stanley}(1971)}]{stanley1971}
\bibinfo{author}{\bibfnamefont{H.~E.} \bibnamefont{Stanley}},
  \emph{\bibinfo{title}{An Introduction to Phase Transitions and Critical
  Phenomena}} (\bibinfo{publisher}{Oxford University Press},
  \bibinfo{address}{Oxford}, \bibinfo{year}{1971}).

\bibitem[{\citenamefont{Goldenfeld}(1992)}]{goldenfeld1992}
\bibinfo{author}{\bibfnamefont{N.}~\bibnamefont{Goldenfeld}},
  \emph{\bibinfo{title}{Lectures on phase transitions and the renormalization
  group}} (\bibinfo{publisher}{Addison-Wesely}, \bibinfo{address}{Reading
  (MA)}, \bibinfo{year}{1992}).

\bibitem[{\citenamefont{Reichl}(1998)}]{reichl1998}
\bibinfo{author}{\bibfnamefont{L.}~\bibnamefont{Reichl}},
  \emph{\bibinfo{title}{A Modern Course in Statistical Physics}}
  (\bibinfo{publisher}{John Wiley \& Sons}, \bibinfo{address}{New York},
  \bibinfo{year}{1998}).

\bibitem[{\citenamefont{James et~al.}(2008)\citenamefont{James, Plank, and
  Brown}}]{james2008}
\bibinfo{author}{\bibfnamefont{A.}~\bibnamefont{James}},
  \bibinfo{author}{\bibfnamefont{M.~J.} \bibnamefont{Plank}}, \bibnamefont{and}
  \bibinfo{author}{\bibfnamefont{R.}~\bibnamefont{Brown}},
  \bibinfo{journal}{Phys. Rev. E} \textbf{\bibinfo{volume}{78}},
  \bibinfo{pages}{051128} (\bibinfo{year}{2008}).

\bibitem[{\citenamefont{Maya et~al.}(2017)\citenamefont{Maya, Miramontes, and
  Boyer}}]{maya2017}
\bibinfo{author}{\bibfnamefont{M.}~\bibnamefont{Maya}},
  \bibinfo{author}{\bibfnamefont{O.}~\bibnamefont{Miramontes}},
  \bibnamefont{and} \bibinfo{author}{\bibfnamefont{D.}~\bibnamefont{Boyer}},
  \bibinfo{journal}{The European Physical Journal Special Topics}
  \textbf{\bibinfo{volume}{226}}, \bibinfo{pages}{391} (\bibinfo{year}{2017}),
  ISSN \bibinfo{issn}{1951-6401},
  \urlprefix\url{https://doi.org/10.1140/epjst/e2016-60195-6}.

\end{thebibliography}

\end{document}